\documentclass[aps,prl,showpacs,twocolumn,groupedaddress]{revtex4}

\usepackage{graphicx,color}
\usepackage{bm, amsmath, amssymb}

\begin{document}

\title{ Quantum Dynamics of a Bose Superfluid Vortex  }

\author{L. Thompson$^{1,2}$ and P.C.E. Stamp$^{1,3}$}

\affiliation{$^{1}$Department of Physics \& Astronomy, University
of British Columbia, Vancouver, BC V6T 1Z1, Canada \\
$^2$Department of Physics,
Massachusetts Institute of Technology,
Cambridge, Massachusetts 02139, USA\\
$^{3}$Pacific Institute of Theoretical Physics, University
of British Columbia, Vancouver, BC V6T
1Z1, Canada}


\begin{abstract}

We derive a fully quantum-mechanical equation of motion for a vortex
in a 2-dimensional Bose superfluid in the temperature regime where
the normal fluid density $\rho_n(T)$ is small. The coupling between
the vortex ``zero mode'' and the quasiparticles has no term linear
in the quasiparticle variables -- the lowest-order coupling is
quadratic. We find that as a function of the dimensionless frequency
$\tilde \Omega = \hbar \Omega/k_BT$, the standard
Hall-Vinen-Iordanskii equations are valid when $\tilde \Omega \ll 1$
(the ``classical regime''), but elsewhere, the equations of motion
become highly retarded, with significant experimental implications
when $\tilde \Omega \gtrsim 1$.

\end{abstract}


\maketitle

Quantum vortices were first predicted in $^4$He superfluid by
Onsager \cite{onsager} and found experimentally a decade later
\cite{vinen}. Vortices, along with quasiparticles, constitute the
two basic excitations in many condensed matter systems
\cite{thoulessB}; they may also have existed as topological defects
in the early universe \cite{earlyU}.

Remarkably, the fundamental question of how vortices and
quasiparticles interact, and how vortices move is very
controversial, notably for superfluids
\cite{TAN96,sonin97,anglin07}. The argument is usually phrased in
terms of forces acting on a vortex moving with respect to both the
superfluid [having local velocity ${\bm v}_s({\bf r})$ and
superfluid density $\rho_s$] and the normal fluid [with velocity ${\bm
v}_n({\bf r})$ and density $\rho_n$]. Then, the classical equation
of motion for the semiclassical vortex coordinate ${\bm R}_v(t)$
(here taken to be a point in the plane -- we discuss the 3D problem
at the end) is usually written \cite{HallV} as
\begin{align}
 \label{eom1}
M_v \ddot {\bf R}_v - {\bm f}_M - {\bm f}_{qp} - {\bf F}_{ac}(t)
\;=\; 0
\end{align}
where ${\bf F}_{ac}(t)$ is some driving force, $M_v$ is the vortex mass,
${\bm f}_M = \rho_s {\bm \kappa} \times (\dot {\bf R}_v - {\bm
v}_s)$ is the (uncontroversial) Magnus force for a vortex with
circulation ${\bm \kappa} =\hat {\bf z}h/m$, and the quasiparticle
force is
\begin{equation}
{\bm f}_{qp} \;=\; D_o ({\bm v}_n - \dot {\bf R}_v) + D_o^{\prime}
\hat {\bf z} \times ({\bm v}_n - \dot {\bf R}_v)
 \label{f-QP}
\end{equation}
where $D_o(T)$, $D_o^{\prime}(T)$ depend strongly on the temperature
$T$. The classic discussion of Iordanskii \cite{iord} yields
\begin{equation}
D_o^{\prime}(T) = - \kappa \rho_n(T)
 \label{iordF}
\end{equation}

These ``HVI equations,'' due to Hall, Vinen, and Iordanskii
\cite{HallV,iord}, have been used to analyze thousands of
experiments in superfluids and superconductors in the last 60 years
\cite{ExptV}. However, there is no consensus on the value of either
the mass $M_v$ (estimates in the literature range from zero to
infinity \cite{anglin07}) or of the coefficients $D_o(T)$,
$D_o^{\prime}(T)$. Indeed, Thouless et. al. \cite{TAN96} find
$D_o^{\prime}(T) = 0$ for all $T$, and scattering analyses give
various results for $D_o^{\prime}(T)$ \cite{sonin97,iord,
fetter64,wexthou98,stone00}. The resolution of these questions has
become an important unsolved problem in physics. We briefly discuss
the experimental situation below.

Previous analyses have been restricted to a local (in space and
time) description, derived from semiclassical or perturbative
calculations of quasiparticle scattering off a static vortex, with
no vortex recoil \cite{sonin97,iord,fetter64,wexthou98,stone00}.
This yields forces acting instantaneously on a quasiclassical vortex
coordinate ${\bf R}_v(t)$. There is no general agreement between
these calculations (which are rendered difficult by the long-range
vortex profile). Our tactic has been to formulate the problem fully
quantum mechanically, in terms of an equation of motion for the
vortex reduced density matrix $\bar {\bm \rho}({\bf R},{\bf R'};t)$
which is obtained by integrating out all quasiparticle degrees of
freedom. Here the vortex coordinate states $|{\bf R} \rangle, |{\bf
R'} \rangle$ are defined by the position of the vortex node
appearing in the many-body wave function. We then define a vortex
``center of mass'' coordinate ${\bf R}_v = {1\over 2}({\bf R}+{\bf
R}')$, and a ``quantum fluctuation coordinate'' ${\bm \xi} = {\bf
R}-{\bf R}'$. Equations of motion are then found for these two
quantum variables, with the vortex recoil automatically incorporated.
Remarkably, in thermal equilibrium we find that the results largely
depend on one key parameter, the ratio $\tilde \Omega = \hbar
\Omega/k_BT$, where $\Omega$ is the characteristic frequency of the
vortex motion, and $k_BT$ is the thermal energy of the
quasiparticles. When $\tilde \Omega \ll 1$ we are in a ``classical
regime,'' where we find that the HVI equations
(\ref{eom1})-(\ref{iordF}) can be justified, with the addition of a
nontrivial fluctuation force on the right-hand side of
(\ref{eom1}). However, when $\tilde \Omega \gg 1$ we are in a quantum
regime which has seen little experimental exploration, and where the
vortex equations of motion are rather different.

Two key features of the analysis \cite{SuppI} are as follows:
(i) The widespread assumption of a vortex-quasiparticle
coupling which is linear in the quasiparticle variables is not
correct. The vortex is a solitonic excitation of the same field as
the quasiparticle excitations. Linear couplings are then forbidden:
for the vortex to be a {\it bona fide} minimum action solution to
the equations of motion, the lowest-order coupling has to be at
least quadratic in all fluctuation variables \cite{raja}. One then
needs to find a ``renormalized'' coupling to new fluctuation variables
that are correctly orthogonalized to each other and to the vortex
``zero mode''; this turns out to be very complicated \cite{SuppI}.
The renormalized coupling, which is singular at low momentum
transfer, is indeed quadratic in these new variables.
(ii) Integration over quasiparticle coordinates then
produces time-retarded, long-range interactions between different
points on a vortex ``worldline.'' A nonperturbative path integral
treatment (required to deal with the singular vortex-quasiparticle
interaction) then yields extra ``memory forces'' in the equations of
motion, which become important in the quantum regime $\tilde \Omega
> 1$.

{\em (i) Results:} At low temperatures, Bose liquids are described
by an effective Hamiltonian of the form \cite{popov}:
\begin{align}
H = \int d^2r \left( {\rho\over 2m_o^2}(\hbar \nabla\Phi)^2 +
\epsilon[\eta, \nabla \eta] \right)
\end{align}
with density  $\rho = \rho_s + \eta({\bf r})$, density fluctuations
$\eta({\bf r})$, an energy functional $\epsilon[\eta, \nabla \eta]$
whose form depends on which superfluid we study, and a superfluid
phase $\Phi({\bf r})$. This Hamiltonian is restricted to
lengthscales $\gg a_o = \hbar/m_o c_o$, to energies $\ll m_oc_o^2$,
and to velocities $\ll c_o$, the sound velocity, where $c_o^2 =
\rho_s (d^2 \epsilon /d \eta^2)|_{\eta = 0}$.

We emphasize the limitations of this hydrodynamic formulation. It is
valid for both low-density Bose gases and dense superfluids like
$^4$He, provided $k_BT \ll m_oc_o^2$ (so that $\rho_n \ll \rho$; in,
e.g., $^4$He superfluid, this means $T \lesssim 0.7$ K), and likewise
for perturbations of frequency $\Omega \ll m_oc_o^2/\hbar$. With
these restrictions it is valid for arbitrary ratios of the
``crossover parameter'' $\tilde \Omega = \hbar \Omega/k_BT$. However
it does not include interquasiparticle interactions, which are very
small in this low-$T$ regime and have no bearing on the form of the
quasiparticle-vortex interaction, but which are essential for the
macroscopic transport of energy and momentum \cite{ThouV}. Nor do we
study here the role of the boundaries -- such geometric effects are
crucial in understanding the vortex mass \cite{anglin07}, although
they hardly affect the vortex-quasiparticle interaction.

(a) {\it Equations of Motion}: The results are more transparent when
Fourier transformed. Defining ${\bf R}_v(t) = \int_0^\infty d\Omega {\bf
R}_v(\Omega) e^{i \Omega t}$, we write the equation of motion as
\begin{equation}
R^v_i(\Omega) = A_{ij}(\Omega, n_{\bf q}) F_j(\Omega),
 \label{eomNew}
\end{equation}
where $n_{\bf q}$ is the quasiparticle distribution over momentum
${\bf q}$, and the total ``driving force''
\begin{equation}
{\bf F} =  {\bf F}_{ac}(\Omega) - q_v {\bm \kappa} \times {\bf
J}(\Omega) + {\bf F}^{fluc}(\Omega, n_{\bf q})
 \label{F-dr}
\end{equation}
sums an external driving field, an internal local transverse force
${\bm f}_{\perp} = - q_v {\bm \kappa} \times {\bf J}(\Omega)$, where
${\bf J} = \rho_s {\bm v}_s + \rho_n {\bm v}_n$ is the total
current, and a fluctuation term ${\bf F}^{fluc}(\Omega, n_{\bf q})$.
The $2 \times 2$ ``admittance matrix'' $A_{ij} = A^{\parallel}
\delta_{ij} + A^{\perp}_{ij} $, where
\begin{eqnarray}
 A^{\parallel}   &=& \mathbb D^{-1} [-\Omega^2
 M_v(\Omega, n_{\bf q})+i \Omega D_{\parallel}(\Omega, n_{\bf q})] \nonumber \\
 A^{\perp}   &=& \mathbb D^{-1}
 [\hat \sigma_y \kappa \rho\Omega - \hat\sigma_x|\Omega| d_{\perp}(\Omega, n_{\bf
 q})]
 \label{A++}
\end{eqnarray}
where $\hat\sigma$ are the usual Pauli matrices, and
\begin{equation}
\mathbb D(\Omega, n_{\bf q}) = [\Omega^2 M_v - i\Omega
D_{\parallel}]^2 - [\kappa\rho \Omega -i|\Omega| d_{\perp}]^2
 \label{D-Omega}
\end{equation}
If we write $A_{ij} = \Omega \sigma_{ij}$, then $\sigma (\Omega,T)$
is analogous to a conductivity tensor, with $D_{\parallel}(\Omega)$
playing the role of the longitudinal resistivity.

The key difference between (\ref{eomNew})-(\ref{D-Omega}) and
previous results for this problem lies in the frequency dependence
of $D_{\parallel}(\Omega, n_{\bf q})$, $d_{\perp}(\Omega, n_{\bf
q})$, $M_v(\Omega, n_{\bf q})$, and in the correlator
$\chi_{ij}(\Omega,n_{\bf q}) = \langle F_i^{fluc} (\Omega,n_{\bf
q})F_j^{fluc}(-\Omega,n_{\bf q})\rangle$ of the fluctuation force
${\bf F}^{fluc}$. Quite generally we find that when the
quasiparticles are in equilibrium at temperature $T$, the
``viscous'' terms [i.e., $D_{\parallel}(\Omega, T)$,
$d_{\perp}(\Omega, T)$, and $\chi_{ij}(\Omega, T)$] in
(\ref{eomNew})-(\ref{D-Omega}) take the form $Q(\Omega, T) = f(T)
g(\tilde \Omega)$. The effective mass $M_v(\Omega, T)$ has a more
complicated behavior.

\begin{figure}
\includegraphics[width=3.2in]{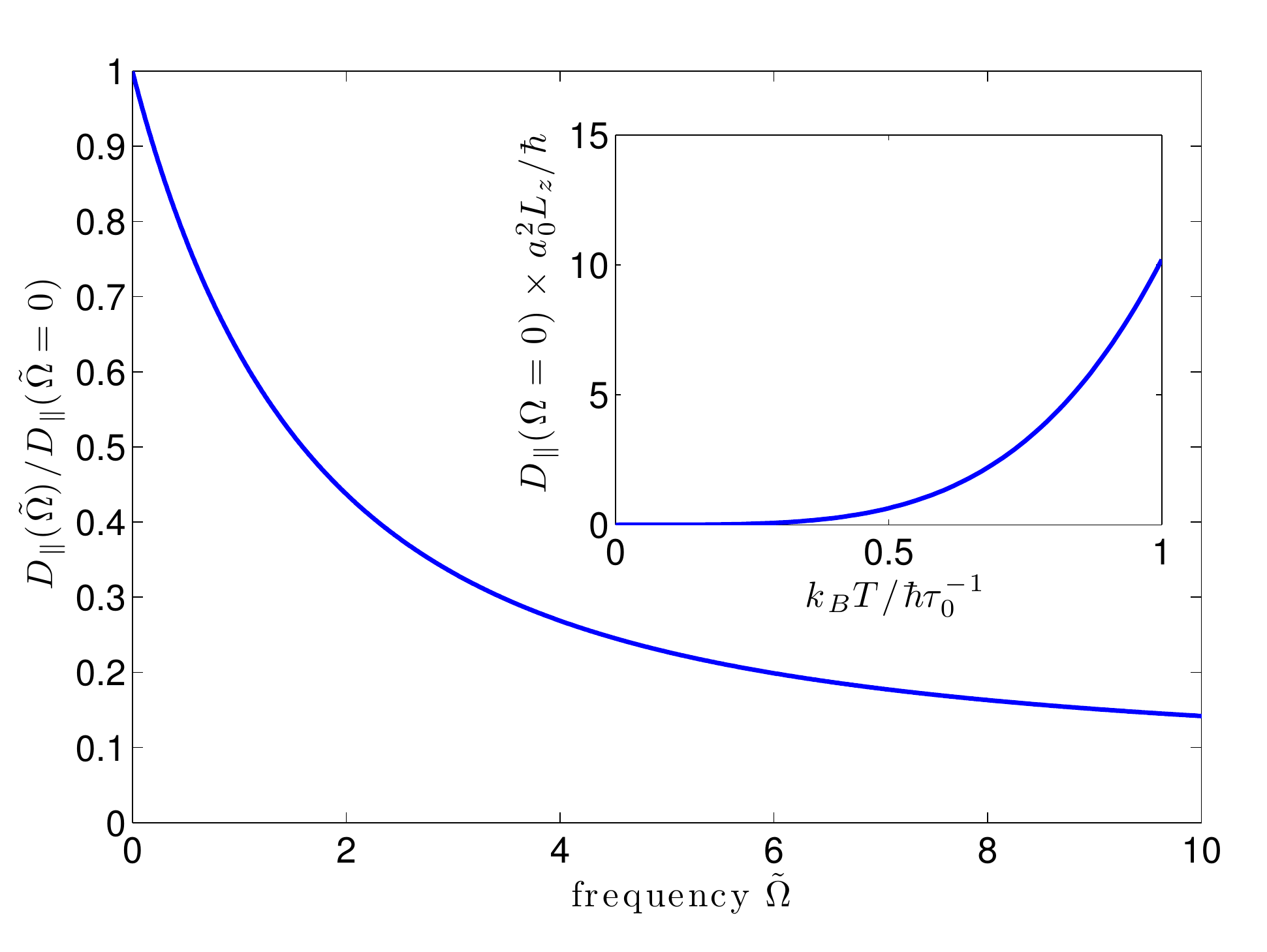}
\caption{\label{fig:Dlong} The longitudinal damping coefficient
$D_\|(\tilde \Omega,T)$. Main figure: The dependence of $D_\|(\tilde
\Omega,T)$ on $\tilde \Omega$, normalized to its zero frequency
value $D_o(T)$. Inset: The coefficient $D_o(T)$, proportional to
$T^4$, plotted as a function of the dimensionless temperature
$(\tau_o/\hbar)k_BT$. }
\end{figure}

\begin{figure}
\includegraphics[width=3.1in]{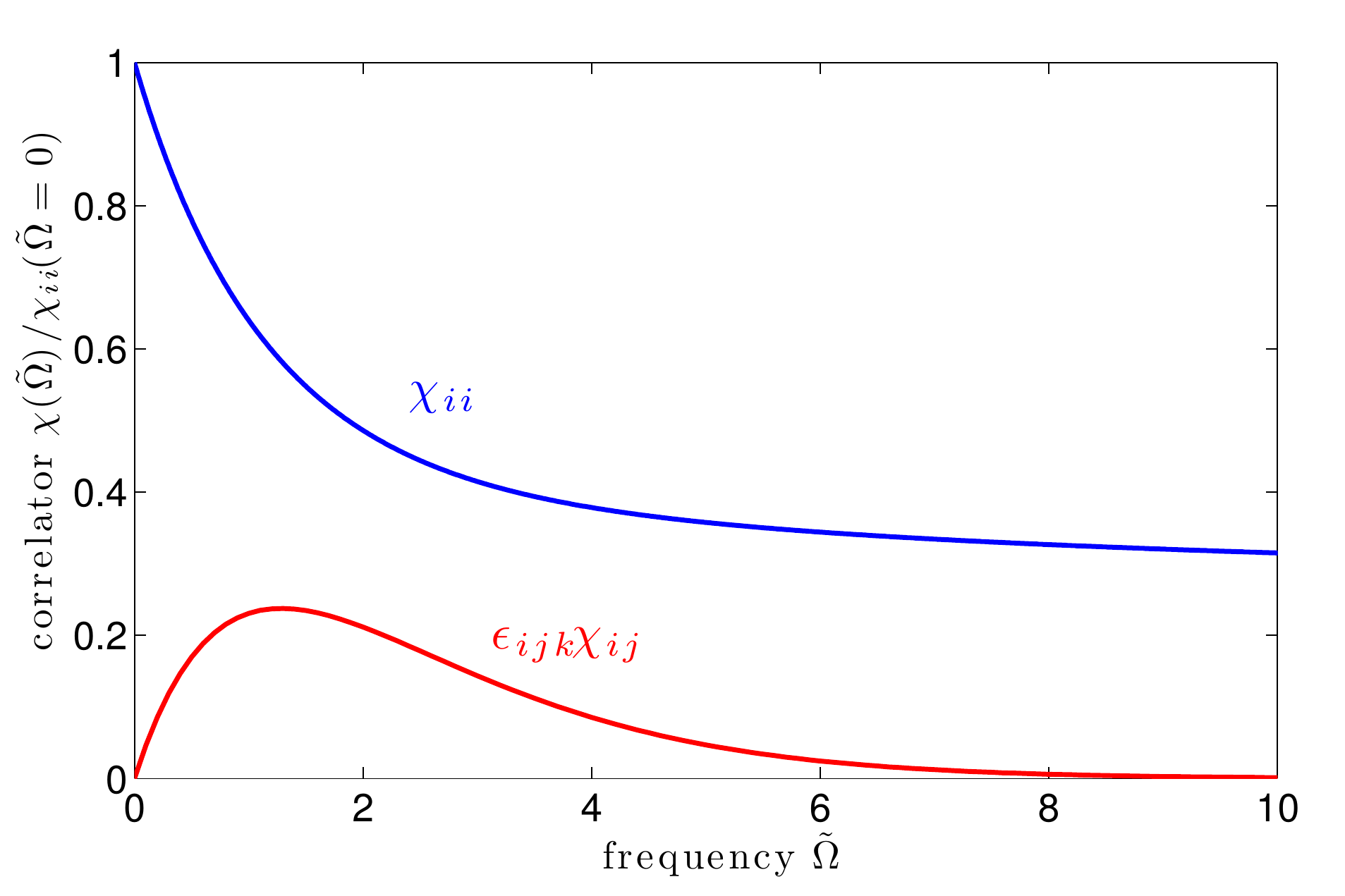}
\includegraphics[width=3.1 in]{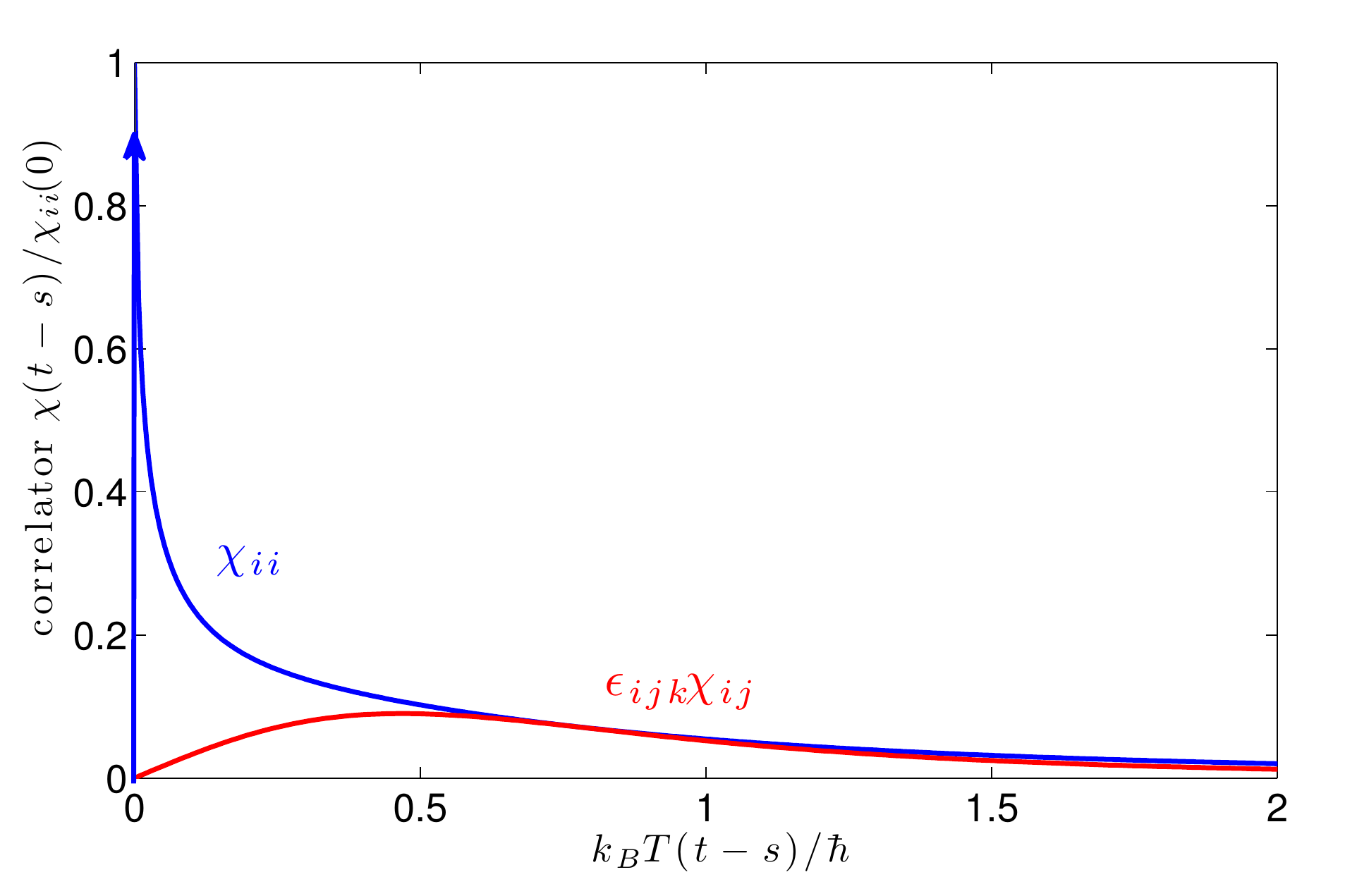}
\caption{\label{fig:Fflcorr} The correlator $\chi_{ij}$ of the
fluctuating force, plotted as a function of $\tilde \Omega$ (top)
and the dimensionless time $k_BT(t-s)/\hbar$ (bottom). The plots are
 normalized to $\chi_{ii}$ at zero frequency (top) and at zero
time (bottom). The arrow at $t-s = 0$ in the bottom figure
represents a $\delta$-function contribution (see text).}
\end{figure}

(b) {\it Quantum-Classical crossover}: Consider first
$D_{\parallel}(\tilde \Omega, T)$, shown in Fig. \ref{fig:Dlong}.
Over the whole range of $\tilde \Omega$ and $T$,
$D_{\parallel}(\tilde \Omega, T) = D_o(T) g_D(\tilde \Omega)$, where
$D_o(T)$ is just the coefficient in (\ref{f-QP}), and where
$g_D(\tilde \Omega)$ decreases smoothly from $g_D(0) = 1$ in the
classical limit to $g_D(\tilde \Omega \rightarrow \infty) = 1/16$ in
the quantum limit \cite{D-asymp}. The fluctuation correlator
$\chi_{ij}(\tilde \Omega, T)$, shown in Fig. \ref{fig:Fflcorr}, is a
little more complicated. Ordinarily, one expects the noise correlator
in a quantum Langevin equation to have the form $\chi_{QL}(\tilde
\Omega, T) \sim f(T) \tilde \Omega \coth {1\over2}\tilde \Omega$,
i.e., a strictly increasing function of $\tilde \Omega$. However Fig.
\ref{fig:Fflcorr}(a) shows a quite different behavior: like
$D_{\parallel}(\tilde \Omega, T)$, the longitudinal correlator
$\chi_{ii}(\tilde \Omega, T)$ {\it decreases} smoothly with $\tilde
\Omega$, now with the limiting behavior
\begin{align}
 \label{FFlongAsymp}
{\chi_{ii}(\tilde \Omega) \over \chi_o^{\parallel}} \rightarrow
\left\{
\begin{array}{ll}
   \left( 1 - {\zeta(3)\over 2\zeta(4)}\tilde\Omega \right)&
  \;\;\;\;\;\; (\tilde \Omega \to 0) \\
{\zeta(5) \over 4\zeta(4)} \; \left( 1 + {5\zeta(6) \over \zeta(5)}
 \; \tilde \Omega^{-1} \right) &  \;\;\;\;\;\;   (\tilde \Omega \to \infty)
\end{array}\right.
\end{align}
where $\chi_o^{\parallel} (T) = \chi_{ii}(\tilde \Omega\!\!=\!\!0,T)=A D_o(T)$, with $A =k_B T /L_z\pi$ (a
relationship coming from the fluctuation-dissipation theorem). On
the other hand, in both limits the transverse part $\epsilon_{ijk}
\chi_{ij}$ of $\chi_{ij} (\Omega, T)$ is zero; it rises to a maximum
value $\sim 0.2\chi_o^{\parallel} (T)$ in the crossover regime
$\tilde \Omega \sim 1$.

A Fourier transform back to the time domain [see Fig.
\ref{fig:Fflcorr}(b)] reveals an initial $\delta$ function in
$\chi_{ii}(t-s,T)$ [because $\chi_{ii}(\tilde \Omega, T)$ is
everywhere finite and positive] followed by a slow decay $\propto
(t-s)^{-2}$. Similar behavior is found for
$D_\|(t-s,T)$ \cite{SuppI}. The transverse correlator rises from zero at zero
time, and decays more rapidly [like $(t-s)^{-3}$] at long times.

The transverse function $d_{\perp}(\tilde \Omega, T)$ also shows a
characteristic crossover behavior -- we do not elaborate here
because $d_{\perp}(\tilde \Omega, T)$ is always very small compared
to $D_{\parallel}(\tilde \Omega, T)$. Full details on all these
functions are found in the Supplemental Material \cite{SuppI}.

Finally, consider the vortex inertial mass $M_v(\Omega, T)$
appearing in (\ref{eomNew})-(\ref{D-Omega}). This is well-known to
depend on the sample geometry \cite{anglin07}; for a circular
container of radius $R_o \gg a_o$, in the $\Omega \rightarrow 0$
limit, we easily verify the well-known hydrodynamic result
\cite{M-hyd}
\begin{equation}
M_v (0) = \pi \rho_s a_o^2 \left[ \ln \left( {R_o \over a_o}\right)
+ \gamma_E + 1/4 \right]
 \label{Mv0}
\end{equation}
where $\gamma_E$ is Euler's constant. Naively, one expects that in
the quantum limit $\tilde \Omega \gg 1$, the $\ln (R_o/a_o)$ factor
in (\ref{Mv0}) will be replaced by $\ln (c_o/a_o \Omega)$ (the
length scale $c_o/a_o$ being set by the distance quasiparticles can
travel in a time $\sim \Omega^{-1}$). However the actual behavior
is more subtle: there is a ``radiation reaction'' term $\propto
d^3{\bf R}_v/dt^3$ in the equation of motion, analogous to that in
electrodynamics, and to deal with this one must go beyond the
expansion in powers of $\dot {\bf R}_v/c_o$ being used here. This
problem lies outside the scope of the present paper \cite{arovas}.

 (c) {\it Real-time dynamics}: Remarkably, the results given above
allow us to write simplified {\it local} equations of motion in {\it
both} quantum and classical limits. In the classical limit, Fourier
transforming back gives precisely the HVI equations
(\ref{eom1})-(\ref{iordF}), but with an added noise term:
\begin{align}
 \label{eom-cl}
M_v \ddot {\bf R}_v - {\bm f}_M - {\bm f}_{qp} - {\bf F}_{ac}(t)
\;=\; {\bf F}_{fluc}^{(cl)}(t)
\end{align}
where the classical noise force ${\bf F}_{fluc}^{(cl)}(t)$ has the
correlator
\begin{equation}
\chi_{ij}^{(cl)}(t-s, T) \sim \chi_o^{\parallel} (T) \delta_{ij}
\delta(t-s)
 \label{chi-cl}
\end{equation}
i.e. an entirely longitudinal Markovian noise. However this equation
is only meaningful on coarse-grained time scales $\gg \hbar/k_BT$; for
shorter times, the time-retarded nature of the correlations becomes
crucial, and as Fig. \ref{fig:Fflcorr} makes clear, the fluctuation
correlator $\chi_{ij}(t-s, T)$ then becomes anisotropic and highly
non-Markovian, and the HVI equations simply do not apply.

In the opposite quantum limit $\tilde \Omega \gg 1$, one may again
write a local equation like (\ref{eom-cl}), of HVI form, again with
an added noisy fluctuating force. However, now the coefficients are
different; $D_o(T)$ in (\ref{f-QP}) is replaced by $D_o(T)/16$, and
the quantum noise correlator $\chi_{ij}^{(Q)}(t-s, T) = {\zeta(5)
\over 4\zeta(4)} \chi_{ij}^{(cl)}(t-s, T)$ (again entirely
longitudinal). In this limit, valid for time scales $\ll \hbar/k_BT$
(but $\gg \hbar/m_oc_o^2$), the coefficients in these ``quantum
HVI'' equations arise solely from the $\delta$-function
contributions to the correlation functions --  all retarded parts
are suppressed.

We would like to emphasize how unusual these results are in detail.
It is quite remarkable to have 2 equations of exactly the same form
(but quite different coefficients) in these two limits, but with a
quite different form in the crossover between them; it is more
illuminating to look at it in frequency space, as above. And yet,
very surprisingly (at least to us), the Iordanskii force is quite
unaffected by this -- apart from the very small correction term
$d_{\perp}(\tilde \Omega, T)$, the Iordanskii force is independent
of frequency, and can be treated as entirely local, and
(\ref{iordF}) is reproduced exactly in our derivation (which is
quite different from previous scattering theory calculations).


%

\vspace{2mm}

\noindent {\bf (ii) Experimental Implications}: The results above justify the
HVI equations \cite{HallV,iord}, and the phenomenology based on
these \cite{ExptV}, in the classical regime. However away from this
regime we find a quite new phenomenology. Clean experimental tests
of this will require (a) that the vortex not be coupled to some
other object (e.g., charged ions), which change its natural dynamics,
and (b) that the vortex be coupled to the natural excitations of the
system (as opposed to, e.g., a source of external quasiparticles,
which can couple linearly to the vortex). The results will also
change in situations where the vortex is being ``dragged'' by some
external time-varying potential \cite{TAN96}, since such potentials
may strongly distort the vortex in the region where they act.

The most obvious direct experimental realization of the results here
would be in 2-dimensional Bose-condensed atomic gases; the dynamics
of single vortices can then be tracked in experiments
\cite{dalibard} (e.g., in their spiraling out from the trap center),
and the viscous coefficients can then in principle be extracted from
such measurements \cite{shlyap}. A detailed treatment using the
present equations is quite lengthy, and will be presented elsewhere.
Our results are also clearly relevant to experiments on turbulence
in superfluid $^4$He, which have recently begun to probe the quantum
regime \cite{golov}, and any theory of vortex tunneling in $^4$He or
cold gases must include the viscous effects described here (which
are very far from being described by a simple Caldeira-Leggett
coupling \cite{CL83}). One difficulty with turbulence or tunneling
is that in most experimental cases the vortices are 3-dimensional
objects, and the vortex line may distort in many ways that are
impossible to capture analytically. It would nevertheless be
interesting to extend, at least numerically, the existing theories
of quantum turbulence \cite{Qturb} and vortex tunneling \cite{vortT}
in $^4$He to include the effects discussed here. Finally, we
emphasize that the results given here are not applicable to
fermionic superfluids like $^3$He or superconductors -- the form of
the vortex-quasiparticle interaction is quite different in these
systems.

\vspace{1mm}

To summarize: within the constraints of the low-$T$ hydrodynamic
picture, we find that the HVI equations can be justified in a purely
quantum-mechanical treatment, with the addition of a fluctuation
noise term, provided one is in the classical regime $\tilde \Omega
\ll 1$. Outside this regime one needs to use a more general set of
equations, which show strong memory effects in the time domain.

\vspace{2mm}

This work was supported by NSERC, CIFAR, PITP, and by the Killam
Trusts. We would like to thank D.J. Thouless and W.G. Unruh for
many useful discussions.


\begin{thebibliography}{99}
\bibitem{D-asymp} We find $D_o(T) \;=\; {3 m_o q_V^2 \pi \zeta(4) \over L_z \tau_o}
        (\tau_o k_BT/ \hbar)^4$ in 2 dimensions. When $\tilde \Omega \ll 1$
        one has $g_D(\tilde \Omega) \sim [1 - {\zeta(3) \over 2\zeta(4)} \; \tilde \Omega]$;
        and for $\tilde \Omega \gg 1$, we get $g_D(\tilde \Omega) \sim {1 \over 16}
        [1 + {12\zeta(5) \over \zeta(4)}\; \tilde \Omega^{-1}]$.
\end{thebibliography}
\end{document}